\documentclass[floatfix,aps,prb,twocolumn,groupedaddress]{revtex4}
\usepackage{graphicx}
\usepackage[]{epsfig}
\usepackage[]{psfig}
\usepackage{times}
\usepackage[abs]{overpic}
\newcommand{\beq}{\begin{equation}}
\newcommand{\eeq}{\end{equation}}
\newcommand{\bea}{\begin{eqnarray}}
\newcommand{\eea}{\end{eqnarray}}
\newcommand{\nn}{\nonumber}

\newcommand{\eps}{\epsilon}

\newcommand{\de}{\delta}
\newcommand{\D}{\Delta}

\newcommand{\app}{\approx}

\begin{document}
\bibliographystyle{apsrev}
\title{Detecting the Kondo screening cloud in conductance measurements
 on quantum dots}
\author{Pascal Simon}
\email{psimon@physics.bu.edu}
\affiliation{Physics Department, Boston University, 590 Commonwealth Ave., 
Boston, MA02215}

\author{Ian Affleck}
\email{affleck@physics.bu.edu}
\altaffiliation{On leave from Canadian Institute for 
Advanced Research and Department of Physics
 and Astronomy, University of British Columbia, Vancouver,
BC,  Canada, V6T 1Z1}
\affiliation{Physics Department, Boston University, 590 Commonwealth Ave., 
Boston, MA02215}
\date{\today}
\begin{abstract}
The observation of the Kondo effect in quantum dots has provided new 
opportunities to finally observe the controversial Kondo screening cloud.
We study how screening cloud effects appear in the conductance through 
a  quantum wire containing a quantum dot when the length of the wire 
is comparable to the size of the screening cloud.
\end{abstract}
\maketitle
One of the most remarkable triumphs of recent progress in nanoelectronics 
has been the observation of the Kondo effect in a single semi-conductor 
quantum dot.\cite{dot,Cronenwett,Wiel}  The quantum dot is formed in a semiconductor heterolayer 
by applying voltages which  confine electrons to an island of size 
a fraction of a micron.  When the tunneling amplitude to the 
dot from the external leads is sufficiently small, the number of electrons 
on the dot becomes quite well-defined leading to the Coulomb blockade 
effect on the conductance.  When the number of electrons on 
the dot is odd, it can behave as an $S=1/2$ magnetic impurity and the 
transmission and back-scattering of electrons is described by a magnetic 
exchange interaction between this impurity and the conduction electrons.  
The Kondo effect is a renormalization of this exchange interaction to 
large values at low temperature, $T<T_K$, where $T_K$ is the 
Kondo temperature, leading ultimately to ideal 
$2e^2/h$ conductance.  This Kondo effect results from the formation 
of a spin singlet between the impurity spin and a conduction electron 
in a very extended wave-function, known as the screening cloud.  The 
size of this screening cloud is of order $\xi_K\approx v_F/T_K$ where 
$v_F$ is the Fermi velocity and is of order 1 micron.

Some of these semiconductor devices have the quantum dot embedded in 
a quantum wire of width comparable to the dot dimension and length 
of order 1 micron.  The ends of the wire must be connected, ultimately, 
to three-dimensional leads in order to perform conductance measurements. 
In this situation, the Kondo screening cloud may fill the entire 
quantum wire and even  extend beyond it into the macroscopic leads.   
One might expect some modification, perhaps suppression, of the Kondo 
effect to occur if the quantum wire is shorter than the screening cloud.
If so, this could provide an ultimate limitation on miniaturization of 
some nanoelectronic devices. 

We have recently investigated this effect in a somewhat different 
situation where the quantum wire containing the quantum dot 
is made into a closed ring.  Unfortunately, it
is then impossible to perform a standard conductance measurement.
However,  the persistent 
current induced by a magnetic flux is also sensitive to screening 
cloud effects and is drastically reduced when the circumference 
of the ring becomes smaller than $\xi_K$.\cite{Affleck}  While this demonstrates 
the importance of screening cloud effects, such a persistent 
current measurement would be a difficult experiment.  Thus 
we turn our attention here to an experimentally easier but theoretically 
more challenging device: a quantum dot embedded in a quantum wire which 
is in turn 
connected to external leads by weak tunnel junctions.    We assume 
that a gate voltage can be applied to the dot and also to the quantum 
wires.  We note that a related device has been proposed recently by Thimm
{\it et al.} \cite{Thimm} where a Kondo impurity was equally coupled to all energy levels
of a finite size  box. 
The energy level spacing was assumed constant and 
of $O(1/V)$ where $V$ is the volume of the (3 dimensional) box. 
These two aspects differ considerably from the geometry studied here,
since the electrons need to pass through the
dot to contribute to the conductance. Moreover, the Non Crossing
Approximation used in [\onlinecite{Thimm}] might be questionable in such
geometry where several new energy scales emerge compared to the usual Kondo model.
\begin{figure}
\epsfig{figure=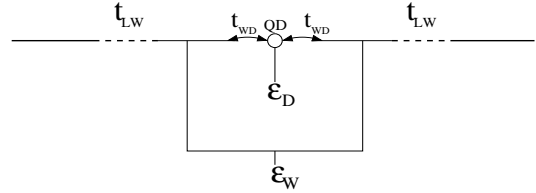,height=2.5cm,width=7cm}
\caption{Schematic representation of the device under consideration. $\eps_D$
  and $\eps_W$ control respectively the dot and wire gate voltage.}
\label{device}
\end{figure}
 A simplified one-dimensional tight-binding model which describes 
our device is indicated in Fig. \ref{device} and has the Hamiltonian:
\begin{equation}
H=H_L+H_W+H_D+H_{LW}+H_{WD},
\end{equation} 
where $L$, $W$ and $D$ stand for leads, wires and dot respectively.  Here:
\begin{eqnarray}
H_L&=&-t\left[ \sum_{j=-\infty}^{-L-2}+\sum_{j=L+1}^{\infty}\right]
(c^\dagger_jc_{j+1}+h.c.)\nonumber \\
H_W&=&-t\left[ \sum_{j=-L}^{-2}+\sum_{j=1}^{L-1}\right]
(c^\dagger_jc_{j+1}+h.c.)\nn   \\&&+\epsilon_W
\left[ \sum_{-L}^{-1}+\sum_{1}^{L}\right]n_j  \\
H_D&=&\epsilon_Dn_0+Un_{0\uparrow}n_{0\downarrow}\nonumber \\
H_{LW}&=&-t_{LW}(c^\dagger_{-L-1}c_{-L}+c^\dagger_{L}c_{L+1}+h.c.)\nonumber \\
H_{WD}&=&-t_{WD}(c^\dagger_{-1}c_0+c^\dagger_0c_1+h.c.)\nn
\end{eqnarray}
Here $n_{j\sigma}\equiv c^\dagger_{j\sigma}c_{j\sigma}$ and 
$n_j\equiv n_{j\uparrow}+n_{j\downarrow}$.
Effects being left out of our simple model include other impurities 
in the quantum wires and the presence of several channels.  (The 
quantum wires studied in [\onlinecite{Wiel}] apparently contained 
about 10 channels.)  We also ignore electron-electron interactions 
in the wires and leads, only keeping them in the dot.  

We will assume that the system is in the strong Coulomb blockade regime, 
so that $t_{WD}<<-\epsilon_D$, $U+\epsilon_D$, where $\epsilon_D<0$.  
Then we may eliminate the empty and doubly occupied states of the 
dot, so that $H_D+H_{WD}$ gets replaced by a Kondo interaction
plus a potential scattering term:
\begin{eqnarray}
H_{WD}+H_D\to H_K&=&J(c^\dagger_{-1}+c^\dagger_1)
{\vec \sigma \over 2}(c_{-1}+c_1)\cdot \vec S \nn\\
&+&V(c^\dagger_{-1}+c^\dagger_1)(c_{-1}+c_1)
\end{eqnarray}
with $J=2t_{WD}^2\left[ {1\over -\epsilon_D}+{1\over U+\epsilon_D}\right]$,
$V={t_{WD}^2\over 2}\left[ {1\over -\epsilon_D}-{1\over U+\epsilon_D}\right]$.
Here $\vec S$ is the spin operator for the quantum dot.  
We assume that the Kondo interaction and the lead-wire tunneling 
are weak, $J, t_{LW}<<t$.

%The Kondo effect can be understood as resulting from a renormalization 
%of the Kondo coupling constant, $J$, to large values at low temperatures.  
%Perturbation theory is infrared divergent but the temperature acts 
%as an infrared cut-off yielding a finite result which is accurate 
%if the temperature is sufficiently high ($T>>T_K$).  At low temperatures, 
%a non-perturbative description is needed.  This is provided by the 
%local Fermi liquid description.\cite{Nozieres}  If we imagine that 
%$J>>t$, then 
%a spin singlet forms in the groundstate from the impurity and an electron 
%in a symmetric orbital on sites $\pm 1$.  The anti-symmetric orbital 
%still remains available to conduct current so the system is roughly 
%equivalent to the $U=0$ model and exhibits resonant conductance with 
%the resonance tied to the Fermi surface \cite{Simon}.  

In the case 
of a closed ring, we showed earlier that the renormalization of the Kondo
coupling 
is cut off, even at low temperatures, by the ring circumference.\cite{Affleck}  In 
the present situation, this renormalization would be cut off by the finite 
length, $L$, of the quantum wires, if $t_{LW}=0$.  Essentially, if the 
Kondo cloud doesn't have sufficient room to form, then the growth 
of the Kondo coupling constant is cut off. The effective 
Kondo coupling at the length scale $L$ becomes of O(1) when 
$L\approx \xi_K$.   

What is less obvious, 
is what happens for small but finite $t_{LW}$.  Then, even if $L<<\xi_K$, 
the Kondo cloud can still form by leaking into the leads.  The growth 
of the Kondo coupling is not cut off by the finite size of the wire.
Nonetheless, we might expect some noticeable effects to occur when 
$L$ is reduced to a value of $O(\xi_K)$, associated with the screening 
cloud beginning to leak into the leads.  It is these effects which we 
wish to study in the present paper.  

It is instructive to begin with a calculation of the conductance 
that treats $t_{LW}$ exactly but treats $J$ in lowest non-vanishing 
order of perturbation theory.  We expect this to be valid 
at sufficiently high $T$ when the renormalized coupling is sufficiently 
small.  To do this calculation, we first diagonalize the Hamiltonian 
at $J=0$. i.e. we diagonalize 
$H_0\equiv H_L+H_W+H_{LW}$.
%  Since the sites 
%at $j\leq 0$ and $j\geq 0$ are completely decoupled in this limit, 
%the spectrum consists of  two-fold degenerate even and odd energy levels 
%(for each electron spin state).  If $t_{LW}=0$, then these wave-functions 
%and eigenvalues are:
%\begin{eqnarray} 
%\psi_{e}(j)&=&(1/\sqrt{L})\sin k|j|\nonumber \\
%\psi_{o}(j)&=&(1/\sqrt{L})\sin kj\nonumber \\
%k&=&\pi n/(L+1)\nonumber \\ 
%E(k)&=& -2t\cos k+\epsilon_W.\end{eqnarray}  
For non-zero $t_{LW}$, the spectrum of $H_0$ is continuous.
In order to study the Kondo interaction in perturbation theory, it is 
useful to express $c_1+c_{-1}$ in terms of the even eigenstates, 
$c_\epsilon$ of $H_0-\mu N_e$:
$
{(c_1+c_{-1})\over \sqrt{2}} = \int_{-2t-\mu}^{2t-\mu} d\epsilon f(\epsilon
)c_\epsilon.
$
We normalize $c_\epsilon$ so that 
$\{ c_\epsilon ,c^\dagger_{\epsilon{'}}\} =\delta (\epsilon -\epsilon{'})$.
 Then $f(\epsilon )$ obeys the normalization condition
$
\int_{-2t-\mu}^{2t-\mu} d\epsilon |f(\epsilon )|^2=1.$
For small $t_{LW}$, this ``local density of states'', 
$\rho (\epsilon )\equiv |f(\epsilon )|^2,$
has sharp peaks at the energies
$
\epsilon_n=\epsilon_W-\mu -2t\cos [k_n],
$
where the momenta  
$k_n\app \pi n/L+O(t_{WL}^2/Lt^2)$. 
The separation between these peaks, near zero energy  is
$\Delta_n \approx \pi v_F/L$.
The 
width of these peaks is approximately
$\delta_n =(2t_{LW}^2\sin^3(k_n)\eta_{n,W}/tL$ with:

$\eta_{n,W}=\sqrt{1+\eps_W\cos k_n/(t\sin^2 k_n)-\eps_W^2/(4t^2\sin^2 k_n)}$.

We see that the ratio of width to separation is of order:
\begin{equation}
\delta_n /\Delta_n \app {t^2_{LW}\over t^2}{\sin^2k_n\over \pi}\eta_{n,W}.\end{equation}
We will assume that this quantity is $<<1$.  
  The full Hamiltonian
may be written in this basis as:
\bea
&&H-\mu N_e=\int_{-2t-\mu}^{2t-\mu} d\epsilon \epsilon
 c^\dagger_\epsilon c_\epsilon\nn\\&& + \int d\epsilon d\epsilon{'}
f^*(\epsilon )f(\epsilon{'})
\left(2Jc^\dagger_\epsilon {\vec \sigma \over 2}c_{\epsilon{'}}
\cdot \vec S+2Vc^\dagger_\epsilon c_{\epsilon{'}}\right)\label{HE}
\eea
The linear conductance (for $t_{LW}\ne 0$) at cubic order in $J,V$  is 
given by:
\bea
G(T) &=&  {e^2\over \pi\hbar}\pi^2\int d\epsilon \rho (\epsilon )^2
[-dn_F/d\epsilon ]{3\over 4}J^2[1+2J~ I_1(\eps)] ,\nn\\
&+&{e^2\over \pi\hbar}4\pi^2V^2\int d\epsilon \rho (\epsilon )^2[-dn_F/d\epsilon ]\label{Gf}
\eea
with $I_1(\eps)=\int d\epsilon' {\rho(\eps')\over
(\epsilon'-\epsilon)}(1-2n_F(\eps'))$. 
$n_F(\epsilon )$ is the Fermi distribution function at temperature T.
Notice that the potential scattering term does not renormalize at this order 
in agreement with [\onlinecite{Glazman}]. The integral $I_1(\eps)$ depends on
the local density of states $\rho(\eps)$.   
 
Let us focus on the second order terms in $J$ and $V$, and
ignore, for the moment, the corrections of higher order. We 
must distinguish 3 regimes of temperature resulting simply from the 
fact that the width of $(-dn_F/d\epsilon)$ is $O(T)$.  If $T>>\Delta_n$, then 
the integral in Eq. (\ref{Gf}) averages over many peaks of 
$\rho (\epsilon )$ so that $G$ is approximately independent of $\epsilon_W$:
\begin{equation}
G\approx {e^2\over \pi \hbar}(\pi \rho_0)^2[3J^2/4+4V^2]\label{GH}
\end{equation}
where $\rho_0=\sin k_F/\pi t$ is the average local density of states.
  When $\delta_n <<T<<\Delta_n$, the conductance 
depends strongly on $\epsilon_W$. 
If $\epsilon_W$ is tuned to a 
resonance peak, $\epsilon_W=\mu+2t\cos[k_n]$, then 
the integral in Eq. (\ref{Gf}) is dominated by the peak at $k_n$ 
and we find:
\begin{equation}
G(T)\approx {e^2\over \pi\hbar}{(3J^2/4 + 4V^2)\over 4TL t_{LW}^2 
\eta_{n,W}}\pi t \sin k_n
%\approx {e^2\over \pi\hbar}(3J^2/4 + 4V^2){v_F^2\D^2\over 16\pi\de_n t^5}
\label{gor}
\end{equation}
On the other hand, if $\epsilon_W$ is far from a resonance peak 
(compared to $T$) then:
\begin{equation}
G(T)\approx {e^2\over \pi\hbar}(3J^2/4+ 4V^2){t_{LW}^4 
\sin^6 k_n \eta_{n,W}^2\over  t^{6}} 
%\approx (3J^2/4+ 4V^2) {\pi^2\de^2 v_F^2\over 4\D^2 t^4}.
\label{Gmo}
\end{equation}

Finally, in the ultra-low temperature regime, $T<<\delta_n$, and
 $\eps_W$ on resonance, we can evaluate:
\begin{equation}
G(T)\approx {e^2\over \pi\hbar}{({3J^2\over4}+4V^2)t^2\over t_{LW}^4
\eta_{n,W}^2\sin^2k_n}
%\approx {e^2\over \pi\hbar}({3J^2\over4}+4V^2)
%{v_F^2\D^2\over 4\pi^2\delta^2t^4}.
\label{Gol}\end{equation}
The conductance is still given by Eq. (\ref{Gmo}) when $\eps_W$ is tuned off 
 resonance for $T<\de_n$. These approximate 
formulas certainly break down when they do not give $G\pi\hbar /e^2<<1$, due to higher 
order corrections in $J$ and $V$.  
So our approximate formulas will certainly break 
down before $T$ is lowered to $\delta$ unless $J<<t^2_{LW}/t$, a 
condition which might typically not be satisfied. When these formulas apply, 
we clearly see that the conductance is much larger when $\eps_W$ is tuned
on resonance. 
 
However there is another, more interesting reason why these formulas 
can break down at low $T$, namely Kondo physics.  The cubic correction 
in Eq. (\ref{Gf}) contains a $\ln T$ term which essentially replaces 
$J$ by its renormalized value at temperature $T$, $J_{eff}(T)$.  We expect 
that this will remain true at higher orders.  
At sufficiently high $T$ we can calculate this quantity to lowest 
order in perturbation theory, using the Hamiltonian in the form of 
Eq. (\ref{HE}).  If the band-width is lowered from $\pm D_0$ 
(where $D_0$ is O($t$)] to $\pm D$, 
then:
\begin{equation}
J\to J+J^2\left[ \int_{-D_0}^{-D}+\int_{D}^{D_0}\right] {d\epsilon \rho (\epsilon )
\over |\epsilon |} \label{RG}\end{equation}
The renormalization of $J$ is quite different depending on how far we 
lower the cut off, $D$.  If $D>>\Delta_n$, the integral in Eq. (\ref{RG}) 
averages over many peaks in the density of states so its detailed 
structure becomes unimportant and we obtain the result for the usual Kondo
model,   
unaffected by the weak tunnel junctions: $J\to J[1+2J\rho_0\ln (D_0/D)]$.

On the other hand, for smaller $D$, 
$D<<\Delta_n$, the renormalization of $J$ in Eq. (\ref{RG}) becomes strongly dependent 
on $\epsilon_W$.  Let us first assume that $\epsilon_W$ is tuned to 
a resonance of the density of states of width $\delta_n$.  
Then the integral in Eq. (\ref{RG}) 
gives a very small contribution as $D$ is lowered from $\Delta_n$ down 
to $\delta_n$ so $J_{eff}(D)$ practically stops renormalizing over 
this energy range.  Finally, when $D<\delta_n$, the density of states 
grows rapidly. By approximating the local density of states by a Lorentzian of 
width $\de_n$,
we can  express the result in terms 
of the change in $J_{eff}(D)$ as $D$ is lowered from $\Delta_n$:
\begin{equation}
J_{eff}(D)\approx J_{eff}(\Delta_n )[1+J_{eff}(\Delta_n ){4\sin^2k_n\over \pi L\delta_n}
\ln ({\delta_n\over D})].\label{jeff}
\end{equation}
The density of states appearing in this renormalization is enhanced by 
a factor of 
$(2t\sin k_n/[L\delta_n])\approx t^2/[t_{LW}^2\eta^2_{n,W}\sin^2 k_n]$.  
This leads to 
a rapid growth of $J_{eff}(T)$. 
On the other hand, if $\epsilon_W$ is 
off-resonance then the density of states is small, of order 
$(\delta_n /\Delta_n^2L)$ so the growth of the Kondo coupling is 
very slow at all energies $D<\Delta_n$. 

Now consider the implications of this renormalization for the value of $T_K$,
defined as the temperature where $J_{eff}(T)$ becomes of $O(1)$.  When
$J_{eff}(T)$ becomes large at $T>>\Delta_n$ then $T_K$ is related to the 
bare Kondo coupling and bandwidth as in the usual case (with no weak links):
$T_K\approx T_K^0\equiv D_0e^{-1/2J\rho_0}$. Furthermore, in this case, 
$T_K$ does not depend strongly on $\epsilon_W$.  We may characterize 
this case by $T_K^0>>\Delta_n$ or equivalently $\xi_K<<L$.  The screening 
cloud fits inside the quantum wires and the weak links do not modify 
the Kondo effect significantly.  

\begin{figure}
\vskip 0.5cm
\begin{overpic}[scale=0.4]{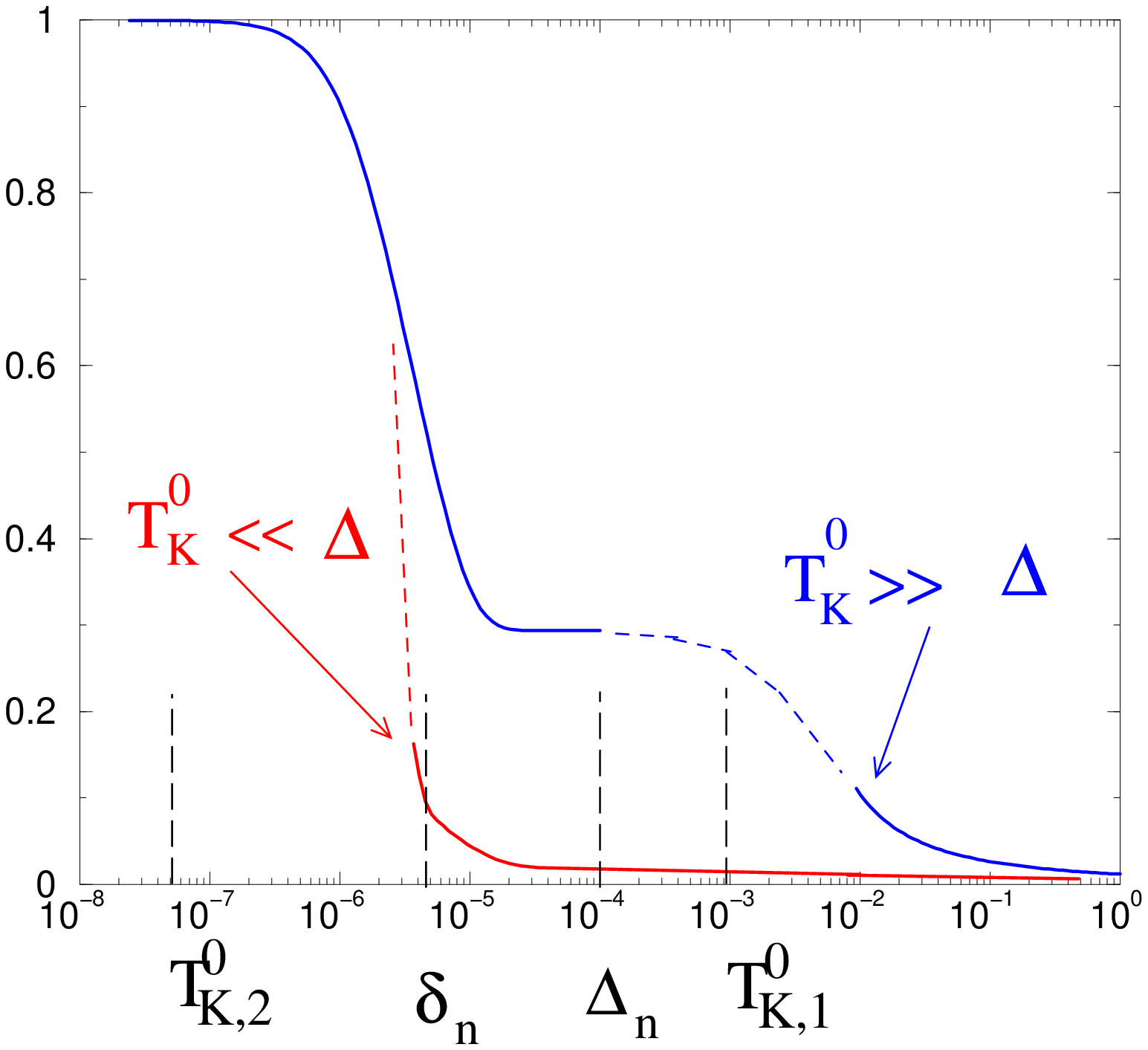}
\put(80,80){\includegraphics[scale=0.17]{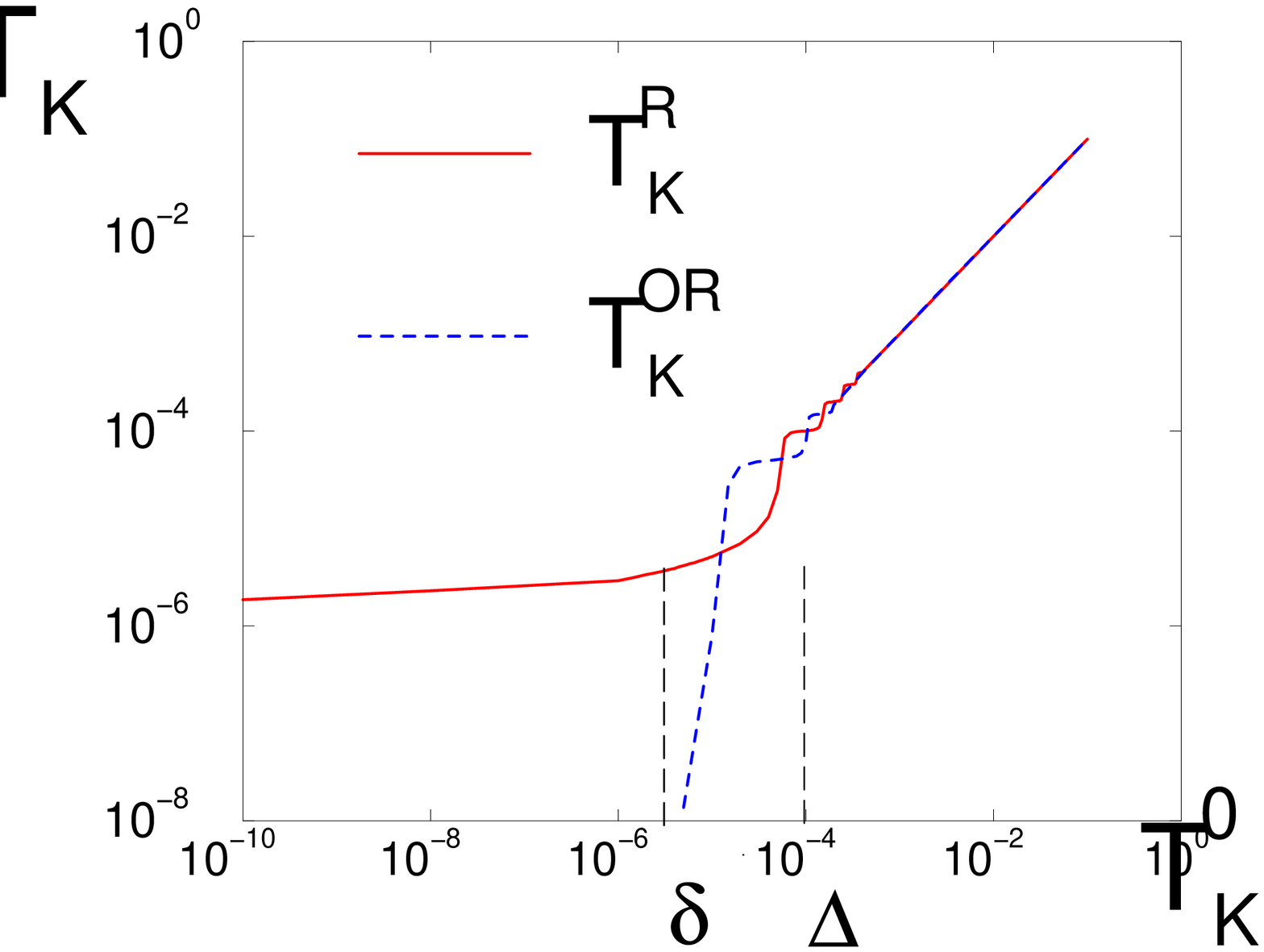}}
\end{overpic}
\caption{Conductance as a funtion  of temperature 
(assuming $\eps_W$ is on resonance) for both cases 
$\D_n\ll T_{K,1}^0$ (right blue
curve) and $\D_n\gg \de_n\gg T_{K,2}^0$ (left  red curve). 
The curves in plain style correspond to the perturbative
calculations plus the Fermi liquid result for the first case only. 
We have 
schematically interpolated these curves (dotted lines) where 
neither the perturbative nor the
Fermi liquid theory applies. The inset represents $T_K=f(T_K^0)$ 
in a log-log scale keeping the same values for $\D_n, \de_n$ for $\eps_W$ on
resonance (plain curve which becomes almost flat at low $T_K^0\ll \de_n$),
and $\eps_W$ off resonance (dashed curve which drops sharly at low $T_K^0$).
Both curves coincide at $T_K^0>\D_n$. 
This clearly illustrates the change of behavior when $\xi_K\gg L$ or 
$\xi_K\ll L$.
}\label{gtres}
\end{figure}

On the other hand, suppose that $T_K^0<<\Delta_n$ implying that
 $J_{eff}(\Delta_n )<<1$.  In this case $T_K$ depends strongly on
 $\epsilon_W$.  If the system is tuned to a resonance then $T_K$ will 
be slightly less than $\delta_n$:
\beq
T_K^{R}\approx \de_n\left( {T^0_K\over
D_0}\right)^{t_{LW}^2\eta_{n,W}^2\sin^2k_n/t^2}=O(\de_n),
\eeq
  On the other hand, if the system 
is off-resonance then $T_K<<\delta_n$:
 \beq
T_K^{OR}\app \Delta_n\left({T_K^0\over D_0}\right)^{t^2/(t_{LW}^2\sin^2k_n)},
\label{tkor}
\eeq 
effectively zero for most purposes.  
This behavior of $T_K$ vs. $T_K^0$ is plotted in the inset of 
Figure \ref{gtres} for both $\eps_W$ on resonance and off
resonance. The curves
coincides for $T_K^0\gg \D_n$ and differ strongly for $T_K^0\ll \D_n$. 
The off resonance Kondo temperature $T_K^{OR}$ drops sharply at $T_K^0<\D_n$
to very small values ($\ll \de_n$). On the other hand $T_K^R$  also has
a sharp drop at $T_K^0<\D_n$ but then becomes almost flat and of order $\de_n$.

Now consider the behavior of the conductance as a function of $T$ and
$\epsilon_W$ in the two cases. In the case $\xi_K<<L$, we may calculate 
the conductance perturbatively in $J_{eff}(T)$ at $T>>T_K^0$ and using 
local Fermi liquid theory for $T<<T_K^0$.  For $T>>T_K^0$, we obtain 
Eq. (\ref{GH}), essentially independent of $\epsilon_W$.  On the 
other hand, for $T<<T_K^0$,  the conductance reduces to that of an ideal wire 
containing no quantum dot, i.e. our original model with $U=0,~
\eps_d=0,~t_{WL}=t$ and some effective length $\tilde L\sim L$. ($\tilde L$
 can be
somewhat reduced from $L$ by an amount of order $\xi_K$). 
%This is given by:
%\beq
%G(T)={e^2\over \pi \hbar}\int 2t\sin k~ dk{4(tt_{LW})^4\sin^2k\sin^2 k'\over
%A^2+B^2}[{-dn_F\over d\eps}(k)]\eeq
%with $k'$ defined by $\cos k'=\cos k-\eps_W/2t$, and
%\bea
%A&=&t^4\sin [k'(2L+2)]-2(t~t_{LW})^2\cos k\sin [k'(2L+1)]\nn\\&+&t_{LW}^4\cos
%2k\sin 2k'L]\nn\\
%B&=&-2(t~t_{LW})^2\sin k\sin [k'(2L+1)]+t_{LW}^4\sin 2k\sin 2k'L\nn
%\eea
As $T$ is lowered below $\Delta_n$ this conductance develops peaks 
with spacing of order $\Delta_n /2$.  
This is the spacing of peaks in the density 
of states of a wire of length $2L$, containing no quantum dot.  It 
is half the spacing in the density of states of the model with $J=0$, 
discussed above.  Initially, as $T$ is lowered below $\Delta_n$, 
the peak width is of $O(T)$ and the peak height is  of 
$O(2e^2\Delta_n t_{WL}^2/hTt^2)$.
As $T$ is lowered below $\delta_n$ the peak width becomes of $O(\delta_n )$
and the peak height becomes of $O(2e^2/h)$.  

On the other hand, when $\xi_K>>L$, the dependence of conductance on $T$ 
and $\epsilon_W$ is very different.  As $T$ is lowered below $\Delta_n$ the 
on-resonance conductance starts to grow both because of the single-electron 
effects reflected in  Eqs. (\ref{gor}) and (\ref{Gol}) and, eventually, 
when $T\leq \delta_n$ because of the growth of $J_{eff}(T)$.  However, 
off resonance the conductance stays small, given by Eq. (\ref{Gmo}) 
 at least down to temperatures, 
$T<<\delta_n$ of $O(T_K^{OR})$, given by Eq. (\ref{tkor}). In the temperature 
regime $T_K^{OR}<<T<<\Delta_n$, the conductance has peaks with spacing 
$\Delta_n$ reflecting the fact that $J_{eff}(T)$ is small, off resonance. 
In this regime it is more difficult to calculate the on-resonance 
conductance both because of the breakdown of the perturbative result of 
Eq. (\ref{gor}), (\ref{Gol})  due to single electron effects and because it appears 
considerably more difficult to extract unambiguous predictions from 
local Fermi liquid theory.  Nonetheless it seems very reasonable to 
expect a conductance of O(1) on resonance at $T\leq \delta_n$   where 
$J_{eff}(T)$ is O(1) on resonance.  Off resonance we can show quite 
rigorously that the conductance remains small since $J_{eff}(T)$ 
remains small there and so do the single electron corrections 
to Eq. (\ref{Gmo}).  Note in particular that the 
values of  $\epsilon_W$ where $T_K$ is large have spacing $\Delta_n$, 
not $\Delta_n /2$.  Thus the halving of the period, which we argued 
above to occur  in the other case, $\xi_K<<L$, does not occur in 
this case at least down to extremely low $T$ of $O(T_K^{OR})$.  (The 
behavior of the conductance 
 at very low $T\leq T_K^{OR}$ in the case $\xi_K>>L$ 
appears more difficult to determine.  However, this is such an 
unphysically low $T$ that it isn't an important limitation of the 
methods that we are using here.)

 \begin{figure}
\epsfig{figure=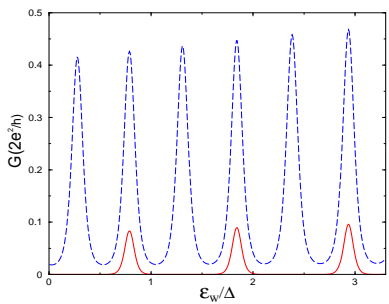,height=5.cm,width=4cm,angle=-90}
\caption{Conductance as a function of $\eps_W$ at fixed $\D_n$, $T$ and $\de_n$
 for both cases $\xi_K\gg L$
(plain style) and $\xi_K \ll L$ (dashed style). We have chosen $T\app \D_n/10\gg
\de_n\app \D_n/100$. The curves have been represented on phase but in general
they are expected to be shifted, the shift being difficult to determine.}
\label{gew}
\end{figure} 

In Figure \ref{gew} we sketch the conductance versus $\epsilon_W$ in these two cases.  
In Figure \ref{gtres}, we have drawn schematically 
the conductance on resonance as a function of temperature for two different bare Kondo
temperatures $T_{K,1}^0\gg \D_n$ and $T_{K,2}^0\ll \de_n\ll\D_n$, using the
perturbative formula given by Eq. (\ref{Gf}) and the Fermi liquid picture 
valid for the first case only.  For the first case, the conductance has  a plateau which corresponds
to the quantum dot being screened and the $\epsilon$ integral in
Eq. (\ref{Gf}) 
 averaging over many peaks.
The conductance reaches $2e^2/h$ only when $T\ll \de_n$. Conversely, in the 
second case,  the conductance
 remains small till $T\approx \delta$
where the Kondo coupling becomes strongly renormalized (see
Eq. (\ref{jeff})).
We may expect a very abrupt increase of the conductance in this regime as
schematically depicted in Fig \ref{gtres}. Notice that for this choice of
$T_{K,2}^0$, the renormalized Kondo temperature $T_{K,2}^R$ is actually
enhanced  and of order $\de_n$. These different behaviors lead to
different shapes of the curves.

So far, we have considered a device which is symmetric around the quantum dot.
This situation might be difficult to reproduce experimentally. One can easily
extend this analysis to the non-symmetric case by considering two local
densities of states $\rho^L$ and $\rho^R$. In this case, by tuning $\eps_W$
 it will be difficult to tune simultaneously $\rho_L$ and $\rho_R$ on
 resonance. 
For example, we can reach the situation when $\rho_L$ is on
 resonance
and $\rho_R$ is not. In the regime $T_K^0\ll \D_n$, this implies that the
renormalized Kondo temperature is controlled by $\rho_L$, meaning that 
the cloud mainly extends into the left wire. 
To have both local density of states on resonance, 
it would  in general be necessary to introduce two independent 
gate voltages $\eps_W^L$ and $\eps_W^R$ controlling the left and right wire.

In conclusion, we have studied how the finite temperature 
conductance and effective Kondo temperature of a quantum dot embedded
in a wire depend  strongly on the ratio between the size of the wire 
and the size of the Kondo screening cloud.

{\bf Acknowledgments} 
We would to acknowledge very helpful discussions with
A. Balseiro, C. Chamon and L. Glazman.

\end{document}